\documentclass[12pt]{article}

\input{epsf.tex}

\newcommand{\E}{{\cal{E}}}
\newcommand{\A}{{\cal{A}}}
\newcommand{\B}{{\cal{B}}}
\newcommand{\C}{{\cal{C}}}
\newcommand{\s}{\sigma}
\newcommand{\I}{{\rm i}}
\renewcommand{\a}{\alpha}

\renewcommand{\d}{{\rm d}}

\newcommand{\be}{\begin{equation}}
\newcommand{\ee}{\end{equation}}
\newcommand{\bea}{\begin{eqnarray}}
\newcommand{\eea}{\end{eqnarray}}

\begin{document}

\title{{\bf On the physical parametrization\\ and magnetic analogs\\ of the Emparan--Teo dihole solution}}
\author{J.~A.~C\'azares$^1$, H.~Garc\'\i a--Compe\'an$^{1,2}$ and
V.~S.~Manko$^1$ \vspace{0.5cm}\\
{\it $^1$Departamento de F\'\i sica,}\\
{\it Centro de Investigaci\'on y de Estudios Avanzados del IPN,}\\
{\it  Apartado Postal 14-740, CP 07000, M\'exico D.F., Mexico}\vspace{0.5cm}\\
{\it $^2$CINVESTAV--IPN, Unidad Monterrey,}\\
{\it Via del Conocimiento  201, Parque de Investigaci\'on}\\ {\it e
Innovaci\'on Tecnol\'ogica (PIIT), Autopista}\\
{\it nueva al Aeropuerto Km. 9.5, Lote 1, manzana 29,}\\
{\it CP. 66600, Apodaca, Nuevo Le\'on, M\'exico}}
\date{}
\maketitle

\abstract{The Emparan--Teo non--extremal black dihole solution is reparametrized using Komar quantities and the separation distance as arbitrary parameters. We show how the potential $A_3$ can be calculated for the magnetic analogs of this solution in the Einstein--Maxwell and Einstein--Maxwell--dilaton theories. We also demonstrate that, similar to the extreme case, the external magnetic field can remove the supporting strut in the non--extremal black dihole too.}

\medskip


\newpage

\section{Introduction}

Several years ago Emparan proposed the name {\it black diholes} \cite{Emp} for the two--body configurations consisting of separated equal black holes carrying opposite electric or magnetic charges. Since then the dihole solutions have been obtained and studies in different non--linear field theories, and one of the approaches to finding new diholes was the use of the known exact solutions of the Einstein--Maxwell equations as starting point in the generation techniques (see, e.g., \cite{LTe}). In the paper \cite{ETe} Emparan and Teo worked out an electrostatic non--extremal dihole solution (henceforth referred to as ETS) belonging to a class of the stationary axisymmetric Einstein--Maxwell fields \cite{MMR}, and then generalized it to the case of the Einstein--Maxwell--dilaton theory and also to the $U(1)^4$ theories arising from compactified field string/M--theory. Among the three arbitrary parameters of ETS only $m$ is the physical mass of each black hole, whereas the remaining parameters $q$ and $k$ do not represent {\it exactly} the physical charge and the coordinate distance between the black--hole constituents. Probably, precisely this fact forced the authors of \cite{ETe} speak about a complicated form of their solution, even though the physical characteristics of the dihole in those parameters looked quite simple.

Fortunately, as was shown in \cite{VCh}, there exists a nice possibility of introducing the individual Komar quantities \cite{Kom} into the multi--black--hole systems via the boundary Riemann--Hilbert problem that simplifies the study of the solutions and makes their physical meaning more transparent; so one of the objectives of our communication is presentation of ETS in terms of the parameters $M$, $Q$, $R$ which are, respectively, the physical mass, physical charge of each black hole and the coordinate distance between the centers of the black hole horizons. This particular purpose will be accomplished by using the results of the recent paper of one of the authors \cite{Man} on the general double--Reissner--Nordstr\"om solution.

Another interesting question related to ETS and called ``a formidable task'' in \cite{ETe} concerns the construction of the potential $A_3$ for the magnetic analogs of ETS. Emparan and Teo found an important relation between magnetic potentials of the Einstein--Maxwell and Einstein--Maxwell-dilaton theories, but did not exploit it, most probably being unaware that Sibgatullin's method \cite{Sib} provides one with a procedure for the calculation of $A_3$ without a need to solve the corresponding system of differential equations. The construction of the potential $A_3$ for the magnetic analog of ETS will be carried out by us with the aid of the integral formulae of paper \cite{RMM}. We show that the exterior magnetic field permits one to achieve equilibrium of the magnetically charged constituents in the non--extremal black dihole.

\section{ETS in physical parameters and its magnetic analog}

A key point in the reparametrization of ETS is the formulae for the irreducible masses of two charged black holes obtained in \cite{VCh}. Since the black holes of ETS are identical and have opposite charges, they possess the same irreducible mass which is defined by the formula \cite{VCh} \be \s=\sqrt{M^2-Q^2+4\kappa MQ}, \quad \kappa\equiv Q/(2M+R), \label{s} \ee where $M$ is the Komar mass of each black hole, $Q$ is the physical charge of the upper constituent (the charge of the lower constituent is $-Q$) and $R$ is the coordinate distance between the centers of the black hole horizons (see Fig.~1).

The knowledge of $\s$ turns out to be sufficient for obtaining the form of the corresponding Ernst potentials \cite{Ern} on the symmetry axis, which in turn allows for working out the expressions of the Ernst potentials and metric functions of ETS in the whole space. For our calculations we used independently the general formulae of $N=2$ Bret\'on--Manko--Aguilar electrostatic solution \cite{BMA} and the expressions defining the double--Reissner--Nordstr\"om solution \cite{Man}. For the Ernst potentials $\E$ and $\Phi$ in both cases we have arrived at the formulae \bea
\E&=&\frac{\A-\B}{\A+\B}, \quad \Phi=\frac{\C}{\A+\B}, \nonumber \\
\A&=&\s^2[(R^2-2M^2+2\kappa^2R^2)(R_++R_-)(r_++r_-) \nonumber\\ &+&4(M^2+\kappa^2 R^2)(R_+R_-+r_+r_-)]  +2[M^4+\kappa^2(4M^4+Q^2R^2) \nonumber\\ &-&4\kappa^3MQR^2](R_+-R_-)(r_+-r_-), \nonumber\\ \B&=&2\s MR(1+4\kappa^2)[\s R(R_++R_-+r_++r_-) \nonumber\\ &-&2M^2(R_+-R_--r_++r_-)], \nonumber\\  \C&=&2\s QR(R-2M)[\s(R_++R_--r_+-r_-) \nonumber\\ &+&2\kappa^2 R(R_+-R_-+r_+-r_-)], \label{EF} \eea where the functions $R_\pm$ and $r_\pm$ have the form \be R_\pm=\sqrt{\rho^2+(z+{\textstyle\frac12}R\pm\s)^2}, \quad r_\pm=\sqrt{\rho^2+(z-{\textstyle\frac12}R\pm\s)^2}.
\label{Rr} \ee Notice, that these functions are defined in a different way than in the paper \cite{ETe}: in our formulae, $r_\pm$ determine the location of the upper constituent and $R_\pm$ of the lower one (see Fig.~1), whereas in \cite{ETe}, for instance, location of the upper black hole is determined by $R_-$ and $r_-$.

For the metric functions $f$ and $\gamma$ which enter Weyl's line element \be
\d s^2=f^{-1}[e^{2\gamma}(\d\rho^2+\d z^2)+\rho^2\d\varphi^2]-f\d
t^2, \label{W} \ee and for the electric potential $A_0$ we then get
\bea
f&=&\frac{\A^2-\B^2+\C^2}{(\A+\B)^2}, \quad e^{2\gamma}=\frac{\A^2-\B^2+\C^2} {K_0 R_+R_-r_+r_-}, \quad A_0=-\frac{\C}{\A+\B}, \nonumber \\
K_0&=&16\s^4R^4(1+4\kappa^2)^2. \label{MF} \eea Formulae (\ref{s})--(\ref{MF}) fully determine the reparametrized ETS.

It is straightforward to check using, for instance, formulae (3.8) of \cite{BMA} that the parameters $M$ and $Q$ ($-Q$) are indeed the Komar mass and charge of the upper (lower) black hole, respectively. The physical meaning of the parameter $Q$ can also be verified with the aid of the simple formula (32) of \cite{ETe} if one takes into account that Emparan and Teo's quantities $m$ and $\kappa_\pm$ are related to our parameters as \be m=M, \quad \kappa_+=R/2, \quad \kappa_-=\s. \label{rel_par} \ee

We conclude the presentation of ETS in terms of physical parameters by observing that the electric dipole moment of the dihole is $Q(R-2M)$, and that the formula (29) of \cite{ETe} for the area of the black hole horizons rewrites as \be A_{bh}=4\pi\frac{(R+2M)^2(\s+M)^2}{R(R+2\s)}, \label{Abh} \ee whence the case of the isolated Reissner--Nordstr\"om black hole is easily recovered in the limit $R\to\infty$.

We now turn to the magnetic analog of ETS within the framework of the Einstein--Maxwell theory. By considering $Q$ ($-Q$) as a {\it magnetic} charge of the upper (lower) black hole, the Ernst potentials  $\E$ and $\Phi$ take the form \be
\E=\frac{\A-\B}{\A+\B}, \quad \Phi=\frac{\I\,\C}{\A+\B}, \label{EF_mag} \ee with the same ${\cal A}$, ${\cal B}$ and ${\cal C}$ as in (\ref{EF_mag}), but the potential $\Phi$ already being a pure imaginary function. The corresponding magnetic potential $A_3$ is defined by the function $\Phi_2$ from the formula (3.13) of \cite{RMM} in the particular $N=2$ case. Then, proceeding in the same way as for the calculation of $\E$ and $\Phi$, and taking into account that $A_3=\Phi_2$ in the magnetostatic case, we finally obtain \bea A_3&=&-\frac{{\cal I}+(z+2M)\C}{\A+\B}, \nonumber\\ {\cal I}&=&Q(R-2M)\{ 4M[\s^2(R_+R_-+r_+r_-)+\kappa^2R^2(R_+r_++R_-r_-)] \nonumber\\ &-&2M(1+\kappa^2)[2M^2(R_+r_-+R_-r_+)+\s R(R_+r_--R_-r_+)] \nonumber\\ &-&R(1+\kappa^2)[\s^2R(R_++R_-+r_++r_-)+6\s M^2 \nonumber\\ &\times&(R_+-R_--r_++r_-)-4M^3(R_++R_--r_+-r_-)] \nonumber\\ &+&4\kappa^2MR^2[R(R_++R_--r_+-r_-)-2\s(R_+-R_-+r_+-r_-)] \nonumber\\ &-&8\s^2MR^2(1+\kappa^2)\}, \label{A3} \eea the magnetic dipole moment of the configuration being $Q(R-2M)$. The metric functions $f$ and $\gamma$ of the magnetic dihole are the same as of ETS, so that the magnetic analog of ETS is completely defined by the above formulae.

The main advantage of having the analytical expression for the potential $A_3$ is a possibility to consider the behavior of a non--extreme black dihole in the external ``uniform'' magnetic field via the Harrison transformation \cite{Har}, as this was done in the extreme case by Emparan \cite{Emp}. Recall that the metric functions $\tilde f$, $\tilde\gamma$ and the magnetic potential $\tilde A_3$ after the action of the Harrison transformation on the known axially symmetric magnetostatic solution $f$, $\gamma$, $A_3$ have the form \bea &&\tilde f=\lambda^2f, \quad e^{2\tilde\gamma}=\lambda^4 e^{2\gamma}, \quad \tilde A_3=2[\lambda^{-1}(1+{\textstyle\frac{1}{2}}BA_3)-1]/B, \nonumber\\ &&\lambda=(1+{\textstyle\frac{1}{2}}BA_3)^2+{\textstyle\frac{1}{4}}B^2\rho^2f^{-1}, \label{Har} \eea where $B$ is a real constant defining the exterior magnetic field.

By acting now with (\ref{Har}) on $f$, $\gamma$ and $A_3$ defined by (\ref{MF}), (\ref{EF}) and (\ref{A3}), we arrive at the solution in which $B$ can be chosen in such a way that the strut between the non--extreme black--hole constituents of the dihole will be removed. Indeed, since $A_3=0$ on the upper and lower parts of the $z$--axis ($\rho=0$, $|z|>{\textstyle\frac{1}{2}}R+\sigma$), the constant $K_0$ in the expression for $\tilde\gamma$ is the same as in (\ref{MF}). On the strut, i.e. on the segment $\rho=0$, $|z|<{\textstyle\frac{1}{2}}R-\sigma$, the potential $A_3$ takes the constant value $2Q$, so that the equilibrium condition $\exp(-\tilde\gamma_0)=1$, where $\tilde\gamma_0$ is the value of $\tilde\gamma$ on the strut, reads as \be \frac{R^2(1+4\kappa^2)}{(R^2-4M^2)(1+BQ)^4} =1, \label{bal} \ee whence for the magnetic field stabilizing the non--extreme magnetic dihole we obtain \be B=\frac{1}{Q}\Biggl(\pm\sqrt[4]{\frac{R^2(1+4\kappa^2)}{R^2-4M^2}}-1 \Biggr). \label{b_eq} \ee Formula (\ref{b_eq}) generalizes Emparan's expression (formula (12) of \cite{Emp}) derived for the extremal Bonnor's dihole \cite{Bon}. The latter expression can be easily recovered from (\ref{b_eq}) if one takes into account that in the extreme case ($\sigma=0$) \be R=2\sqrt{M^2+a^2}, \quad Q^2=M^2r_+^2/a^2, \label{relation} \ee where $r_+$ and $a$ are the parameters employed in the paper \cite{Emp} (the mass parameter $M$ is the same for both cases).

\section{Dilatonic magnetic dihole}

In the case of the Einstein--Maxwell--dilaton theory (see, e.g., \cite{GMa,GHS}) which arises from the Lagrangian \be {\cal L}={1\over 16\pi}\sqrt{-g}\,\,[R-2(\nabla\phi)^2-e^{-2\a\phi}F^2], \label{L} \ee where $\phi$ is the dilaton field and $\a$ the coupling constant ($\a=\sqrt{3}$ for the Kaluza--Klein theory, $\a=1$ for the low
energy effective limit of string theory and $\a=0$ for the pure
Einstein--Maxwell fields), the solution--generating procedure developed in \cite{ETe} consists in the following.

Let $f$, $\gamma$ and $A_3$ be a known magnetostatic solution of the Einstein--Maxwell equations. Then the corresponding magnetostatic solution ($\widehat f$, $\widehat\gamma$, $\widehat{A}_3$, $\phi$)  of the Einstein--Maxwell--dilaton theory is given by the formulae \bea \widehat f&=&f^{\frac{1}{1+\a^2}} e^{-2\a\phi_0}, \quad \widehat\gamma=\frac{1}{1+\a^2}\gamma+\gamma_0, \nonumber\\ \widehat A_3&=&\frac{1}{\sqrt{1+\a^2}}A_3, \quad e^{-2\phi}=f^{\frac{\a}{1+\a^2}} e^{2\phi_0}, \label{GF} \eea where $\widehat f$ and $\widehat\gamma$ are the metric coefficients in the line element \be
\d s^2=\widehat f^{-1}[e^{2\widehat\gamma}(\d\rho^2+\d z^2)+\rho^2\d\varphi^2]-\widehat f\,\d
t^2, \label{mD} \ee $\widehat{A}_3$ is the magnetic potential, $\phi_0$ an arbitrary harmonic function and $\gamma_0$ is obtainable from $\phi_0$ in quadratures via \be \d\gamma=(1+\a^2)[\rho(\phi_{0,\rho}^2-\phi_{0,z}^2)\d\rho +2\rho\phi_{0,\rho}\phi_{0,z}\d z]. \label{g} \ee

The application of the procedure (\ref{GF}) to the magnetic dihole solution considered in the previous section leads to the dilatonic magnetic dihole of the form \bea \widehat f&=&\biggl[\frac{\A^2-\B^2+\C^2} {(\A+\B)^2}\biggr]^{\frac{1}{1+\a^2}}e^{-2\a\phi_0}, \nonumber\\ e^{2\widehat\gamma}&=&\biggl[\frac{\A^2-\B^2+\C^2} {K_0R_+R_-r_+r_-}\biggr]^{\frac{1}{1+\a^2}}e^{2\gamma_0},  \nonumber\\ \widehat{A}_3&=&-\frac{1}{\sqrt{1+\a^2}}\, \frac{{\cal I}+(z+2M)\C}{\A+\B}, \label{MFD} \eea where $\A$, $\B$, $\C$, ${\cal I}$ are the same as in (\ref{EF}), (\ref{A3}), and \bea e^{2\phi_0}&=&\biggl[\biggl(\frac{R_++R_--2\s}{R_++R_-+2\s}\biggr) \biggl(\frac{r_++r_--2\s}{r_++r_-+2\s}\biggr)\biggr]^{-\frac{\a}{1+\a^2}}, \nonumber\\ e^{2\gamma_0}&=& \biggl[\frac{R_+R_-+\rho^2+(z+{\textstyle\frac12}R)^2-\s^2}{2R_+R_-} \nonumber\\ &\times&\frac{r_+r_-+\rho^2+(z-{\textstyle\frac12}R)^2-\s^2}{2r_+r_-} \nonumber\\ &\times&\frac{R_+r_-+\rho^2+z^2-({\textstyle\frac12}R+\s)^2}{R_+r_++\rho^2+(z+\s)^2-{\textstyle\frac14}R^2} \nonumber\\ &\times&\frac{R_-r_++\rho^2+z^2-({\textstyle\frac12}R-\s)^2}{R_-r_-+\rho^2+(z-\s)^2-{\textstyle\frac14}R^2} \biggr]^{\frac{\a^2}{1+\a^2}}. \label{f0} \eea Note that the choice of $\phi_0$ and $\gamma_0$ in (\ref{f0}) is the same as in the paper \cite{ETe}, so we simply give these functions using a different parametrization and different notations for $R_\pm$ and $r_\pm$. Formulae (\ref{MFD}), (\ref{f0}), (\ref{s}), (\ref{EF}) and (\ref{A3}) entirely define the magnetic analog of the dilatonic ETS, and one only has to take into account that due to the presence of the dilaton field the exact physical interpretation of the parameters $M$ and $Q$ slightly changes; for instance, the magnetic charge of the upper black--hole constituent is equal to $Q/(1+\a)^{1/2}$. Obviously, the dilatonic magnetic dihole exhibits essentially the same physical properties as its electric counterpart obtained and analyzed in \cite{ETe}.

\section{Conclusion}

Quite surprisingly, the introduction of the Komar quantities into ETS have not complicated the form of that dihole solution but, on the contrary, only simplified it, making specific physical characteristics of the two--body configuration more visible. So we anticipate that this could make the dihole solutions accessible to a wider physical audience, and not exclusively to experts acquainted with solution--generating techniques. We have also shown that the non--extremal magnetic dihole solutions do not have any serious technical problem of finding the corresponding magnetic potential since Sibgatullin's method gives a straightforward procedure of its calculation; therefore, the choice of the magnetic or electric dihole solution in a particular application now becomes purely a matter of the scientific context of the problem to be considered. Like in the case of the extremal Bonnor--type dihole, the external magnetic field is able to stabilize the non--extremal dihole constituents by regularizing the part of the axis which separates them. Lastly, we mention that our results can be also extended to the non--extreme dihole solutions of the $U(1)^4$ model coming from string/M--theory and reported in \cite{ETe}.

\subsection*{Acknowledgements}

This work was partially supported by Projects 45713--F and 45946--F from Consejo Nacional de Ciencia y Tecnolog\'\i a, Mexico.

\begin{figure}[htb]
\centerline{\epsfysize=60mm\epsffile{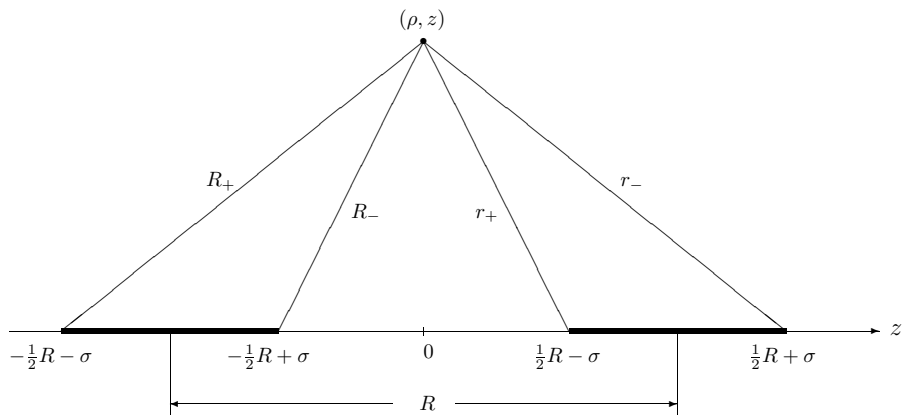}} \caption{Location of the black dihole on the symmetry axis and interpretation of $R_\pm$, $r_\pm$, $R$ and $\s$. The upper constituent is the one related to $r_\pm$.}
\end{figure}

\end{document}